\documentclass{Interspeech}

\interspeechcameraready

\title{Non-Intrusive Binaural Speech Intelligibility Prediction Using Mamba for Hearing-Impaired Listeners}

\author[affiliation={1}]{Katsuhiko}{Yamamoto}
\author[affiliation={1}]{Koichi}{Miyazaki}

\affiliation{}{CyberAgent}{Japan}
\email{\{yamamoto\_katsuhiko, miyazaki\_koichi\_xa\}@cyberagent.co.jp}

\usepackage{comment}

\begin{document}

\maketitle

\begin{abstract}
    
    Speech intelligibility prediction (SIP) models have been used as objective metrics to assess intelligibility for hearing-impaired (HI) listeners. In the Clarity Prediction Challenge 2 (CPC2), non-intrusive binaural SIP models based on transformers showed high prediction accuracy. However, the self-attention mechanism theoretically incurs high computational and memory costs, making it a bottleneck for low-latency, power-efficient devices. This may also degrade the temporal processing of binaural SIPs. Therefore, we propose Mamba-based SIP models instead of transformers for the temporal processing blocks. Experimental results show that our proposed SIP model achieves competitive performance compared to the baseline while maintaining a relatively small number of parameters. Our analysis suggests that the SIP model based on bidirectional Mamba effectively captures contextual and spatial speech information from binaural signals. 
\end{abstract}

\section{Introduction}
\label{sec:introduction}

Speech intelligibility (SI) is an index used to evaluate how spoken words and sentences are heard under noisy and distorted conditions. 
This helps assess the listening abilities of hearing-impaired (HI) individuals and the effectiveness of speech enhancement (SE) processes in hearing aids (HAs) \cite{dillon_hearing_2012,loizou_speech_2013,falk_objective_2015}.
Many SI prediction (SIP) models have been suggested for modeling speech perception and developing objective evaluation tools \cite{relano-iborra_speech_2022,feng_nonintrusive_2022,karbasi_asr-based_2022}. 
However, establishing a unified framework for comparing SIP models has been challenging due to the difficulty in collecting extensive subjective evaluation data.

To address this issue, the Clarity Project\footnote{\url{https://claritychallenge.org/}} includes the Clarity Enhancement Challenge (CEC) and the Clarity Prediction Challenge (CPC) \cite{graetzer_clarity-2021_2021,akeroyd_2nd_2023,barker_1st_2022,barker_2nd_2024}.
As a representative method in CPC2, a non-intrusive binaural SIP model (\texttt{E011})\cite{cuervo_temporal-hierarchical_2023} using the transformer \cite{vaswani_attention_2017} architecture has been proposed, with ``speech foundation model (SFM)'' features extracted from Whisper \cite{radford_robust_2022} and WavLM \cite{chen_wavlm_2022}. 
It outperformed the binaural model of the hearing-aid speech perception index (HASPI) version 2 (\texttt{beHASPI}) \cite{kates_hearing-aid_2021} and all other metrics \cite{barker_2nd_2024}. 

\begin{figure*}[htbp]
    \centering
    \includegraphics[scale=0.8]{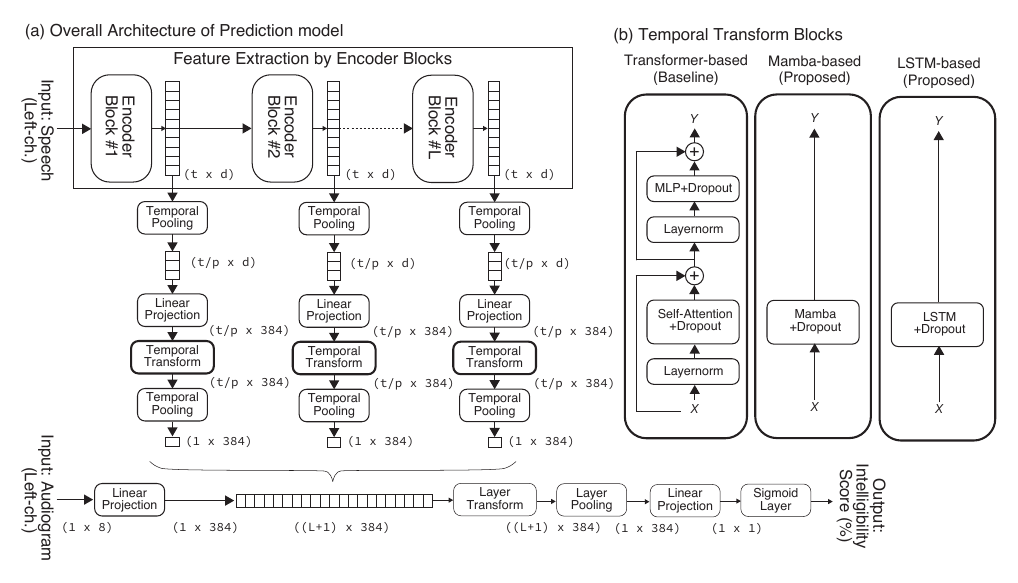}
    \vspace{-5pt}
    \caption{(a) Overview of the monaural SIP model based on \texttt{E011} \cite{cuervo_temporal-hierarchical_2023}. The numbers in parentheses indicate the size of the matrix, where $L$ is the number of encoder layers in Whisper, $t$ is the number of frames obtained as the output of the encoder layer, $p$ is the temporal pooling size, and $d$ is the feature dimension. (b) Architectures of the temporal transform blocks. Input $X$ is the feature of a channel, and output $Y$ is the feature obtained from the output of the blocks.}
    \label{fig:overview}
\end{figure*}

As exemplified by the \texttt{E011} model in CPC2, transformer-based methods have been used de facto in non-intrusive binaural SIP models \cite{tu_unsupervised_2022,zezario_mbi-net_2022,cuervo_speech_2024,mogridge_non-intrusive_2024,zezario_non-intrusive_2024}.
However, the self-attention mechanism theoretically requires $\mathcal{O}(L^2)$ computations and significant memory.
This requirement creates a bottleneck in embedded environments such as HAs, where long-duration input, ultra-low latency, and power efficiency are essential.
By contrast, state-space sequence models (SSMs), such as Mamba, can perform inference in linear time with constant memory \cite{gu_2024_mamba}.
This capability allows them to maintain long-range dependencies while operating in real time, making them suitable for low-computational and low-power devices.

Therefore, in this paper, we propose a Manba-based SIP model to address the limitations of transformer-based models.
Instead of using a temporal transformer, our model leverages Mamba, which is effective for time-series data, including speech applications, with lower inference costs \cite{gu_2024_mamba,miyazaki_exploring_2024}.
Using monaural and binaural models, we compared the prediction performance of SI between transformer- and Mamba-based models. 
Additionally, we further investigated the impact of reducing the temporal pooling size of Whisper features on time resolution and how it affects SI in the Mamba-based model. 

The contributions of this study are as follows:
\begin{itemize}
    \item Our proposed model, featuring a Mamba-based temporal transform block, performed competitively in SI prediction compared to the transformer-based non-intrusive binaural SIP model \cite{cuervo_temporal-hierarchical_2023}.
    \item The experimental results show that our proposed model captures temporal information in Whisper features to perform contextual and binaural processing using Mamba 
\end{itemize}

\section{Mamba}

The Mamba architecture \cite{gu_2024_mamba} was proposed as a sequence-to-sequence model for mapping an input sequence $\mathbf{x} \in \mathbb{R}^D$ to $\mathbf{y} \in \mathbb{R}^D$ with an element-wise hidden state $\mathbf{h}_d \in \mathbb{R}^N$, which can be formulated as follows: \cite{miyazaki_exploring_2024}: 
\begin{align}
    h_{t,d} &= \bar{\mathbf{A}} h_{t-1,d} + \bar{\mathbf{B}} x_{t,d}, \\
    y_{t,d} &= \mathbf{C} h_{t,d},  \\
    \bar{\mathbf{A}}, \bar{\mathbf{B}} &= \exp(\Delta \mathbf{A}), \Delta \mathbf{B}, 
\end{align}
where $\mathbf{A} \in \mathbb{R}^{N \times N}$, $\mathbf{B} \in \mathbb{R}^{N \times 1}$, $\mathbf{C} \in \mathbb{R}^{1 \times N}$ and $\Delta \in \mathbb{R}_+$ represent continuous SSM parameters \cite{gu_parameterization_2022}. 
The Mamba approach proposes a selective scan mechanism to derive input-dependent parameters with $\mathbf{B}_t$, $\mathbf{C}_t$ and $\Delta_t$ as follows: 
\begin{align}
    \mathbf{B}_t, \mathbf{C}_t &= \mathbf{W}_B \mathbf{x_t}, (\mathbf{W}_C \mathbf{x_t})^{\top}, \\
    \Delta_t &= \text{softplus}(\mathbf{W}_{\Delta} \mathbf{x_t}),
\end{align}
where $\mathbf{W}_B \in \mathbb{R}^{N \times D}$, $\mathbf{W}_C \in \mathbb{R}^{N \times D}$, $\mathbf{W}_{\Delta} \in \mathbb{R}^{1 \times D}$, and $(\cdot)^{\top}$ denotes the transpose. 
Mamba also uses a parallel scan \cite{smith_simplified_2023,blelloch_prex_1990} and hardware-efficient algorithms for input-dependent SSM to mitigate the limitations of global convolution \cite{gu_parameterization_2022}. 
The Mamba block combines the prior SSM architecture (H3) \cite{fu_hungry_2022} with a multi-layer perceptron (MLP) block, as detailed in \cite{gu_2024_mamba}. 

\section{Proposed Model}
\label{sec:proposed}

\subsection{Monaural SIP Model}
\label{ssec:mono_sip}

\subsubsection{Overall Architecture}

Figure~\ref{fig:overview}(a) shows the overall architecture of the monaural version of the \texttt{E011} model \cite{cuervo_temporal-hierarchical_2023}.
The model is based on the network structure of a method called Whisper-AT \cite{gong_whisper-at_2023}, which was proposed to perform speech recognition and environmental sound classification jointly.
The model inputs are a single-channel speech signal and audiogram data.
The audiogram represents the listener's hearing threshold in an ear, measured for subjective evaluation experiments at various frequencies \cite{barker_1st_2022,barker_2nd_2024}.

\textbf{Speech Feature Extraction Blocks:} 
The encoder layers of Whisper are used as the backbone for feature extraction.
We also utilized the encoders of the Whisper Large-v2 model, which consist of 1280-dimensional features and $32$ transformer layers.
The parameters of the base model are fixed at their pre-trained values.
Note that the input speech signal is temporarily downsampled to $16$ kHz.
Feature sequences from Whisper encoders are reduced in the time direction by applying an average pooling size $p = 20$ to decrease computational cost \cite{cuervo_temporal-hierarchical_2023,cuervo_speech_2024}.

\textbf{Temporal Transform Block:} 
The Whisper feature sequence is linearly projected into a $384$-dimensional space. 
Then, a $384$-dimensional temporal transform block described in Sec.~\ref{ssec:temporal_transform_mono} is applied to each layer to create a sequence with contextual features.
Global average pooling is applied along the time axis to each layer's sequence, producing a $384$-dimensional embedding for each layer.
These embeddings are concatenated into an $L \times 384$-dimensional matrix.

\textbf{Audiogram Features:} 
The listener's audiogram information is linearly projected onto a $384$-dimensional space and then concatenated with a layer-directional embedding representation, incorporating the listener's characteristics \cite{cuervo_temporal-hierarchical_2023}.

\textbf{Layer-wise SIP Block:} Each feature is concatenated and used as sequence data for layer-wise representation, resulting in an $(L+1) \times 384$-dimensional matrix.
This matrix is input into the single-head layer-directional transformer and then compressed again by global average pooling, yielding a $384$-dimensional representation.
The representation is linearly projected into a single value to generate the SI.
Finally, the output of the sigmoid layer is multiplied by $100$ to produce a prediction result between $0\%$ and $100\%$.

\subsubsection{Implementation of Temporal Transform Block}
\label{ssec:temporal_transform_mono}

Figure~\ref{fig:overview}(b) shows the model architectures of the temporal transform block in Figure~\ref{fig:overview}(a).
The transformer-based architecture is based on the original binaural model of \texttt{E011} \cite{cuervo_temporal-hierarchical_2023} and is modified for monaural processing.
A $384$-dimensional single-head attention module is used for the self-attention block.
We add a skip connection in the MLP block to prevent gradient loss.

We propose a Mamba-based temporal transform block, including a $384$-dimensional Mamba and dropout \cite{srivastava_dropout_2014} module.
In this study, the Mamba module is defined as a unidirectional or bidirectional model because the SSM has a causal operation.
This can be formulated as: 
\begin{align}
    y_{\mathrm{uni}} &= \text{Mamba}(x_t), \\
    y_{\mathrm{bi}} &= \text{Bi-Mamba}(x_t) \nonumber \\
        &= \text{Mamba}(x_t) + \mathrm{flip}(\text{Mamba}(\mathrm{flip}(x_t))), 
\end{align}
where the $\mathrm{flip}$ operation reverses the order of time-series data $x_t$. 
We also considered LSTM-based models, including a $384$-dimensional unidirectional or bidirectional long short-term memory (LSTM) \cite{hochreiter1997long} with a linear projection and dropout module, for comparison with Mamba. 

Among these models, the transformer-based model in Figure~\ref{fig:overview}(b) is used for the layer transform block in Figure~\ref{fig:overview}(a) because temporal information is embedded into a single $384$-dimensional feature in each layer at the layer-wise SIP stage. 

\begin{figure}[tbp]
    \centering
    \includegraphics[scale=0.7]{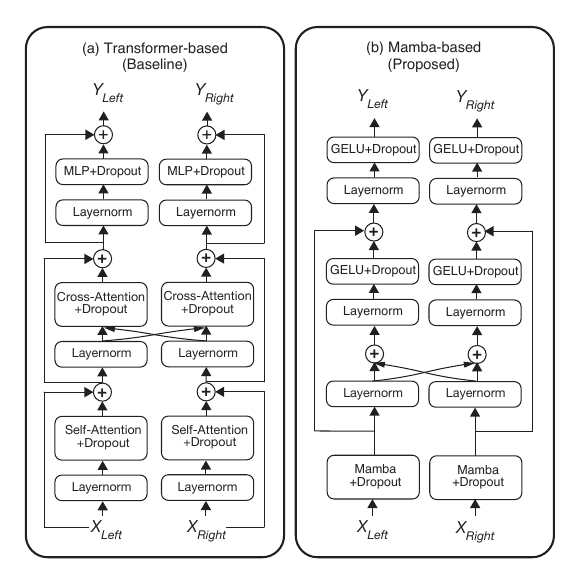}
    \vspace{-10pt}
    \caption{Overview of the (a) transformer-based and (b) Mamba-based binaural temporal transform blocks. The inputs $X_{*}$ are the features of the left and right channels and the outputs $Y_{*}$  are the features obtained from the output of the blocks.}
    \label{fig:binaural_models}
\end{figure}

\subsection{Binaural SIP Model}
\label{ssec:binaural_sip}

The architecture of the binaural SIP model is the same as that of the original \texttt{E011} model \cite{cuervo_temporal-hierarchical_2023}.
The differences from the monaural models in Figure~\ref{fig:overview}(a) are as follows: (1) inputs are the left and right channels of speech features and audiograms, (2) temporal and layer transform blocks are extended for binaural processing, and (3) the latent features of both channels are averaged after the layer pooling block to calculate an SI value.

Figure~\ref{fig:binaural_models} shows the binaural version of temporal transform blocks.
The transformer-based model includes a $384$-dimensional single-head cross-attention block for binaural processing.
This network enables nonlinear interactions using the interaural time and intensity differences between each channel, enhancing binaural SIP accuracy \cite{cuervo_speech_2024}.
For the Mamba-based binaural model, we also propose a binaural processing part with a skip connection and Gaussian error linear units (GELU) after the Mamba module to account for binaural latent features without a cross-attention module.

For the same reason as in the monaural model, the transformer-based model in Figure~\ref{fig:binaural_models}(a) was used for the layer transform block in both binaural models.

\section{Experimental Setups}

\subsection{Code Implementations for Models}
We used code from the GitHub repository for CPC2\footnote{\url{https://github.com/claritychallenge/clarity}} to train and test the baseline and our proposed models.
To implement the baseline model, we used the public code of Whisper-AT \cite{gong_whisper-at_2023}\footnote{\href{https://github.com/YuanGongND/whisper-at}{\url{https://github.com/YuanGongND/whisper-at}}}, allowing feature extraction from encoder layers of Whisper.
We employed the official Mamba codebase and parameters\footnote{\href{https://github.com/state-spaces/mamba}{\url{https://github.com/state-spaces/mamba}}}. 
In this study, dropout modules with a rate of $0.1$ are applied to the self-attention, cross-attention, and MLP layers in the monaural and binaural transformer-based transform blocks.
We set dropout modules to $0.3$ in the Mamba- and LSTM-based transform blocks to prevent overfitting. 

\begin{table}[tbp]
    \centering
    \caption{Overview of CPC2 dataset.}
    \label{tab:dataset}
    \vspace{-5pt}
    \scalebox{0.80}[0.80]{
    \begin{tabular}{cccc}
         \hline
         \textbf{Partition} & \textbf{\#systems} & \textbf{\#listeners} & \textbf{\#samples} \\
         \hline
         \texttt{CEC2.train.1}&  7&  10& 2779\\
         \texttt{+CEC1.train.1}&  10&  22& 5124\\
         \hline
         \texttt{CEC2.test.1}&  3&  5& 305\\
         \hline
         \texttt{CEC2.train.2}&  7&  10& 2796\\
         \texttt{+CEC1.train.2}&  10&  23& 5539\\
         \hline
         \texttt{CEC2.test.2}&  3&  5& 294\\
         \hline
         \texttt{CEC2.train.3}&  7&  10& 2772\\
         \texttt{+CEC1.train.3}&  10&  24& 5820\\
         \hline
         \texttt{CEC2.test.3}&  3&  5& 298\\
         \hline
    \end{tabular}
    }
\end{table}

\subsection{Datasets and Conditions}

The CPC2 dataset\footnote{\url{https://claritychallenge.org/docs/cpc2/cpc2_download}} was used in the subsequent experiments.
This dataset includes speech sentences enhanced by binaural HA processing under simulated noisy conditions.
As shown in Table~\ref{tab:dataset}, the dataset contains the results of subjective SI evaluations of ten types of HA systems and 15 experimental participants (listeners) for CEC2 \cite{akeroyd_2nd_2023}.
We used different partitions of the training data for validation.
For example, when \texttt{CEC2.train.1} and \texttt{CEC1.train.1} were used as the training data, \texttt{CEC2.train.2} and \texttt{CEC1.train.2} were used as the validation data.
All Whisper features were statistically normalized based on each \texttt{train} dataset.

All the models were trained to minimize the Huber loss over 80,000 steps using Adam \cite{kingma_adam_2015} as the optimization method, with a learning rate of $3 \times 10^{-5}$, $\beta_1 = 0.90$, $\beta_2 = 0.98$, and a batch size of 160.
We employed a cosine learning rate scheduler with a linear warm-up over 2,000 steps.
All hyperparameters were set according to the baseline model \cite{cuervo_temporal-hierarchical_2023}.

\begin{table*}[tbp]
\centering
\caption{Prediction results of monaural SIP models based on transformer, LSTM, and Mamba. The mean values of the root-mean-squared error (RMSE) and normalized cross-correlation coefficient (NCC) were calculated from the results of the three trials.}
\label{tab:result_test_monaural}
\scalebox{1.00}[1.00]{
    \begin{tabular}[m]{ccccccc} \hline 
     \textbf{Method} & \textbf{Temporal}& \textbf{\#params} & \multicolumn{4}{c}{\textbf{RMSE($\downarrow$) / NCC($\uparrow$)}} \\ 
     \cline{4-7}& \textbf{Transform Block}& \textbf{(M)} &  \textbf{\textit{CEC2.test.1}} &  \textbf{\textit{CEC2.test.2}} &  \textbf{\textit{CEC2.test.3}} & \textbf{\textit{Average}} \\
     \hline
                    & \multicolumn{1}{l}{Transformer} & 4.05 & 31.65 / 0.68 & 31.24 / 0.68 & 25.46 / 0.80 & 29.45 / 0.72\\
        Baseline   & \multicolumn{1}{l}{\hspace{5pt}- skip at self-attention}& 4.05& 32.15 / 0.67& 29.80 / 0.69& 25.84 / 0.79&29.26 / 0.72\\
                    & \multicolumn{1}{l}{\hspace{5pt}- MLP Block} & 2.86& 33.25 / 0.64& 32.26 / 0.66& 25.98 / 0.80 & 30.50 / 0.70\\
     \hline
                & \multicolumn{1}{l}{(Uni-)LSTM} & 3.46 & 32.42 / 0.67 & 32.16 / 0.67 & 26.04 / 0.79 & 30.21 / 0.71\\
                & \multicolumn{1}{l}{(Bi-)LSTM} & 4.94 & 33.66 / 0.65 & 32.09 / 0.67 & 25.01 / 0.81 & 30.25 / 0.71 \\ 
    \cline{2-7}
                & \multicolumn{1}{l}{(Uni-)Mamba} & 3.24& \textbf{31.02 / 0.67}&  30.65 / 0.68& 23.08 / 0.83& 28.25 / 0.73\\
        Proposed     & \multicolumn{1}{l}{\hspace{5pt}+ skip at Mamba} & 3.24 & 32.80 / 0.66 &  33.46 / 0.65 &  26.81 / 0.78 &  31.02 / 0.70 \\
                & \multicolumn{1}{l}{\hspace{5pt}+ MLP Block} & 4.42 &  31.68 / 0.67 &  29.22 / 0.71 & \textbf{22.76 / 0.83} &  27.89 / 0.74 \\
     \cline{2-7}
                & \multicolumn{1}{l}{(Bi-)Mamba} & 4.20& 31.91 / 0.66& 30.30 / 0.69& 23.24 / 0.83&28.48 / 0.73\\
           & \multicolumn{1}{l}{\hspace{5pt}+skip at Mamba} & 4.20 & 32.71 / 0.65 & 31.52 / 0.67 & 24.83 / 0.81 & 29.69 / 0.71\\
                & \multicolumn{1}{l}{\hspace{5pt}+ MLP Block} & 5.38 &  31.48 / 0.67 &  \textbf{28.47 / 0.72} & 23.69 / 0.82 & \textbf{27.88 / 0.74}\\
    \hline
   \end{tabular}
}
\end{table*}

\begin{table*}[tbp]
\centering
\caption{Prediction results using transformer- and Mamba-based binaural SIP models. The mean values of RMSE and NCC were calculated from the results of 15 trials.}
\label{tab:result_test_binaural}
\scalebox{1.00}[1.00]{
    \begin{tabular}[m]{ccccccc} \hline 
     \textbf{Method} & \textbf{Temporal}& \textbf{\#params} & \multicolumn{4}{c}{\textbf{RMSE($\downarrow$) / NCC($\uparrow$)}} \\ 
     \cline{4-7}
     & \textbf{Transform Block}& \textbf{(M)} &  \textbf{\textit{CEC2.test.1}} &  \textbf{\textit{CEC2.test.2}} &  \textbf{\textit{CEC2.test.3}} & \textbf{\textit{Average}} \\
     \hline
      Baseline  & \multicolumn{1}{l}{Transformer}& 5.23 & 30.05 / 0.69 & 29.26 / 0.71 & 23.16 / 0.83 & 27.49 / 0.74 \\
     \hline
        & \multicolumn{1}{l}{(Uni-)Mamba}& 3.53 & 30.43 / 0.69 & 29.69 / 0.71 & 23.49 / 0.83 & 27.87 / 0.75 \\
      Proposed  & \multicolumn{1}{l}{\hspace{5pt}+ MLP Block} & 4.72 & 30.48 / 0.70 & 30.41 / 0.71 & 23.06 / 0.84 & 27.98 / 0.75 \\
     \cline{2-7}
      & \multicolumn{1}{l}{(Bi-)Mamba}& 5.01 & \textbf{30.03 / 0.70} & 29.11 / 0.72 & \textbf{22.88 / 0.84} & \textbf{27.34 / 0.75}\\
        & \multicolumn{1}{l}{\hspace{5pt}+ MLP Block} & 6.27 & 30.49 / 0.69 & \textbf{28.96 / 0.72} & 23.26 / 0.83 & 27.57 / 0.75 \\      
    \hline
   \end{tabular}
}
\end{table*}

\section{Experiments and Results}
\label{ssec:result}

We conducted comparative monaural and binaural SIP experiments. 
In the monaural experiments, we used the left channel of input speech data. 
We compared the root-mean-squared error (RMSE) and normalized cross-correlation coefficient (NCC) between the predicted and subjective SI results. 

\subsection{Monaural SIP models}

Table~\ref{tab:result_test_monaural} shows the comparative evaluation results using test data (\texttt{CEC2.test.1-3}) with the monaural version of the baseline (\texttt{E011}) \cite{cuervo_temporal-hierarchical_2023} and proposed models.
The mean values of the RMSE and NCC were calculated from the results of three trials.

Transformer-based models predicted SI with an average RMSE of around 29\%.
As in the ablation studies, we removed the skip connection in the self-attention block and the MLP block in the temporal transform block.
The RMSEs of these models were marginally higher or lower than those of the original model.

Our proposed models using unidirectional Mamba reduced the average RMSE by more than 1\% compared with the baseline model.
We conducted additional experiments with both the skip connection and the MLP block, and the results showed that the proposed models using unidirectional and bidirectional Mamba with the MLP block reduced the overall average RMSE.
However, this was not a fair comparison regarding the number of parameters; therefore, we used the simple Mamba block for subsequent experiments.
By contrast, the LSTM-based models did not improve the SI prediction accuracy of the baseline model. 

\subsection{Binaural SIP models}

Table~\ref{tab:result_test_binaural} shows the comparative evaluation results using test data (\texttt{CEC2.test.1-3}) with the binaural version of the baseline and proposed models.
The baseline model with self- and cross-attention transformers in the temporal transform block in Figure~\ref{fig:binaural_models}(a) exhibits lower RMSEs than the monaural version.

Our proposed model, using the binaural model with bidirectional Mamba in the temporal transform block in Figure~\ref{fig:binaural_models}(b), showed the lowest average RMSEs in \texttt{CEC2.test.1}, \texttt{CEC2.test.3}, and the overall average despite having fewer learnable parameters than the baseline model\footnote{The Wilcoxon signed-rank test was performed based on the methodology described in \cite{cuervo_speech_2024}, and no significant differences were found between the models.}.
Surprisingly, the binaural model using the bidirectional Mamba significantly improved over the monaural model, whereas the unidirectional model showed only small improvements.
This indicates that the bidirectional Mamba model effectively embeds contextual and spatial information related to the direction of the sound sources and head rotations in each channel to predict the binaural SI of the CPC2 dataset.

We also considered replacing the final layer normalization, GELU, and Dropout modules in our proposed Mamba-based model with an MLP block.
However, this model predicted SI with a higher average RMSE than the original model.  

\subsection{Effects of Temporal Resolution on Monaural SIPs}

Finally, we investigated the effect of the temporal resolution of input Whisper features on monaural SIPs. 
In this experiment, the temporal average pooling size $p$ in Figure~\ref{fig:overview}(a) was changed from $p = 20$ \cite{cuervo_temporal-hierarchical_2023,cuervo_speech_2024} to $10$ and $5$, where a lower value corresponds to a higher temporal resolution of the original Whisper features. 

Table.~\ref{tab:rmse_pooling} shows the monaural SIP results by baseline and our proposed models. 
The RMSEs did not significantly change, even when the temporal pooling size $p$ of the input features was reduced. 
We expected that by using long time series processing, the Mamba model would effectively capture high-resolution temporal information in Whisper features; however, this was not the case. 
These results imply that temporal information, such as the temporal fine structure component for speech perception in noisy and binaural hearing \cite{moore_roles_2019}, is widely distributed among Whisper features. 

\begin{table}
    \centering
\caption{Relationship between the average RMSE of the monaural SIP model and temporal pooling size of input features. The RMSEs were calculated from the results of three trials.}
\label{tab:rmse_pooling}
    \begin{tabular}{ccccc}
         \hline
         \textbf{Method}&  \textbf{Temporal}&  \multicolumn{3}{c}{\textbf{Pooling size $p$}}\\
         \cline{3-5}
         &  \textbf{Transform Block}&  \textbf{\textit{20}}&  \textbf{\textit{10}}& \textbf{\textit{5}}\\
         \hline
         Baseline&  Transformer&  29.45&  29.36& 30.64\\
         \hline
         Proposed&  (Uni-)Mamba&  \textbf{28.25}&  \textbf{28.53}& 28.62\\
         \cline{2-5}
         &  (Bi-)Mamba&  28.48&  28.97& \textbf{28.22}\\
         \hline
    \end{tabular}
    
\end{table}

\section{Conclusions}

In this paper, we propose a Mamba-based non-intrusive binaural SIP model for HI listeners.
Our proposed model exhibits competitive performance compared with the transformer-based model under both monaural and binaural conditions, confirming the effectiveness of Mamba in temporal transform blocks.
Our analysis implies that the bidirectional Mamba-based SIP model captures contextual and spatial speech information extracted from binaural HA outputs. In future work, we aim to compare the behaviors of the Mamba-based and transformer-based models. 

\bibliographystyle{IEEEtran}
\bibliography{references}

\end{document}